\documentclass[%
 prb,
 reprint,
superscriptaddress,
 amsmath,amssymb,
 aps,
]{revtex4-2}

\usepackage{physics}
\usepackage{graphicx}% Include figure files
\usepackage{dcolumn}% Align table columns on decimal point
\usepackage{bm}% bold math
\usepackage{hyperref}% add hypertext capabilities
\usepackage{verbatim}   %needed for comments
\usepackage[dvipsnames]{xcolor}    %needed for comments
\usepackage[normalem]{ulem}  %needed for strikethrough font
\usepackage{placeins} %for floatbarrier
\usepackage{makecell} %multiline in tables
\usepackage{mathtools}
\begin{document}

\preprint{APS/123-QED}

\title{Enhanced quantum sensitivity and coherence of symmetric magnetic clusters}%
\author{Lorenzo Amato}
\affiliation{PSI Center for Scientific Computing, Theory and Data, CH-5232 Villigen PSI, Switzerland}
\affiliation{Laboratory for Solid-State Physics, ETH Zürich, CH-8093 Zürich, Switzerland}
\author{Manuel Grimm}
\affiliation{PSI Center for Scientific Computing, Theory and Data, CH-5232 Villigen PSI, Switzerland}
\affiliation{Laboratory for Solid-State Physics, ETH Zürich, CH-8093 Zürich, Switzerland}
\author{Markus Müller}
\affiliation{PSI Center for Scientific Computing, Theory and Data, CH-5232 Villigen PSI, Switzerland}

\date{\today}% 

\begin{abstract}

{The search for highly coherent degrees of freedom in noisy solid-state environments is a major challenge in condensed matter. In disordered dipolar systems, such as magnetically doped insulators, compact clusters of two-level systems (TLS) have recently been shown to have significantly longer coherence times than typical single TLS. Coupling weakly to their environment, they sense and probe its many-body dynamics through the induced dephasing. However, it has remained an open question whether further mechanisms exist that protect the coherence of such solid-state qubits. Here we show that symmetric clusters of few TLS couple even more weakly to their surroundings, making them highly sensitive quantum sensors of slow many-body dynamics. Furthermore, we explore their use as qubits for quantum information storage, detailing the techniques required for their preparation and manipulation. Our findings elucidate the role of symmetry in enhancing quantum coherence in disordered and noisy systems, opening a route toward a sensitive experimental probe of many-body quasi-localization dynamics as well as the development of quantum technologies in solid-state systems.}

\end{abstract}

\maketitle

\section{Introduction}

It has been a long-standing goal in condensed matter physics and quantum engineering to host coherent quantum bits (qubits) in a solid-state platform. On the one hand, such qubits are of interest for the purpose of computation or information storage, where decoupling from the environment is key~\cite{bertaina2007rare,grimm2021universal,gonzalez2021scaling,loss1998quantum}. On the other hand, such objects may also serve as quantum sensors for the many-body dynamics of their environment~\cite{davis2023probing,beckert2024emergence,PRXQuantum.3.040328}, which they probe via the weak dephasing caused by the dynamical noise.

For instance, such coherent qubits were discovered in randomly doped magnets, where they appeared in the form of close pairs of ions, whose excitations are particularly well protected from decoherence~\cite{beckert2024emergence}. These objects can sense the flipping of the surrounding majority spins under the dipolar quantum dynamics. The natural question arose as to whether there might be further, even more effective mechanisms that protect coherence. This would facilitate the encoding of quantum information within an environment of similar ensembles of noisy and long-range interacting two-level systems, enabling both increased sensitivity in sensing applications and longer coherence times for quantum information processing and storage.

In this paper, we discuss such mechanisms. One of them relies on the spatial symmetry of small ion clusters, which protects the degeneracy of certain cluster excitations, leading to a substantial reduction of the dominant decay and decoherence channels. In earlier works the related idea of encoding qubits in the lowest states of antiferromagnetic spin clusters had been proposed~\cite{PhysRevLett.104.200502,PhysRevB.86.161409,PhysRevLett.94.207208}. However, such an encoding suffers from dephasing by magnetic field gradients. In contrast, our proposal is based on magnetically neutral excitations that delocalize over small clusters. Those are insensitive to static magnetic field gradients. 

{A second protection mechanism that suppresses decoherence applies to certain states in clusters with an even number of ions. We will show that an effective particle-hole symmetry protects quantum states at the center of the excitation spectrum from a part of the environmental noise.}

Such cluster states allow for fast encoding and manipulation of quantum information, without compromising their lifetime and coherence. This renders these clusters interesting for quantum sensing as well as for information storage.

{
The paper is structured as follows: In 
Sec.~\ref{sec:review} we review the physics of random dipolar magnets and their many-body dynamics that may be sensed by the clusters we discuss in the remainder of the paper. In
Sec.~\ref{sec:model} we specify the required properties of magnetic ions, and discuss the structure and characteristics of quantum states on clusters of ions.
In Sec.~\ref{sec:coherence} we analyze different decoherence channels and show how to counteract them by dynamical decoupling. We then discuss how the remaining dephasing senses the many-body dynamics, and analyze the additional protection of states in the middle of the cluster spectrum. Sec.~\ref{sec:driving} presents techniques to manipulate qubits encoded in clusters. Finally, in Sec.~\ref{sec:conclusions} we discuss the importance of our results and give an outlook on experimental developments.
}

\section{Many-body dynamics and quasi-localization in dipolar magnets}
\label{sec:review}
As mentioned in the introduction, particularly coherent and well-protected qubits may be used as quantum sensors of the dynamics occurring in their surrounding. The latter can be {exceptionally} slow, in which case its detection requires very weakly coupled sensors with long decoherence times.
A class of systems exhibiting particularly interesting, and often very slow, many-body dynamics consists in random ensembles of dipolar interacting two-level systems (TLS)~\cite{Burin1994TheLE,PhysRevB.91.094202,PhysRevB.92.104428,PhysRevB.93.245427,PhysRevLett.113.243002,burin2006energy}. Such two-level systems can be of magnetic, electric or elastic nature and occur ubiquitously in imperfect solid-state materials. Their marginally long-range dipolar coupling helps to overcome the disorder-induced localization tendency, which leads to non-trivial dynamical behavior.
While genuine many-body localization~\cite{nandkishore2015many} (i.e., the lack of long-range transport and thermalization despite interactions) presumably only exists rigorously in short-range interacting one-dimensional systems of discrete degrees of freedom~\cite{imbrie2016many, de2024absence}, signatures of disorder-induced slow many-body dynamics are much more common, being present also in high-dimensional and/or long-range interacting systems. 

For power-law interacting systems, such as dipoles, it has been predicted that, as the strength of disorder is varied, a crossover occurs in the dominant mechanism that  allows for thermalization and transport: The single-excitation hopping regime~\cite{levitov1999critical,PhysRevLett.64.547}, dominating at moderate disorder, eventually should give way to a strong disorder regime where interaction-assisted ``spectral diffusion"  facilitates the dynamics~\cite{PhysRevB.93.245427,burin2006energy,PhysRevLett.113.243002, gornyi2017spectral}. 

Common realizations of such systems are given by magnetic ions doped into an insulating matrix. At low temperatures, such ions can indeed often be described as a randomly distributed ensemble of TLSs subject to dipolar interactions.
{In Ref.~\cite{beckert2024emergence}, such a random magnet, $\text{LiYF}_4$ doped with rare-earth (RE) ions $\text{Tb}^{3+}$, was studied with electron paramagnetic resonance (EPR). The $\text{Tb}^{3+}$ ions are non-Kramers ions, which feature singlet ground and first excited states, well separated from higher energy levels of the $4f$ shell. While the levels of these non-Kramers ions still acquire magnetic moments through the coupling to their nuclear spins, a fraction of them can be rendered insensitive to magnetic noise (to first order) by subjecting them to a so-called ``clock-state condition"~\cite{PhysRevLett.95.060502}, where an external longitudinal magnetic field $B_z$ exactly cancels the hyperfine coupling to their nuclear spin (provided that it has the specific spin projection $I_z$ matched by $B_z$).

The EPR experiment of Ref.~\cite{beckert2024emergence} measured the Hahn echo~\cite{PhysRev.80.580} of such clock-state ions of different energies. Among those, surprisingly coherent degrees of freedom were found to be associated with atypically compact ion pairs. Due to their internal dipolar interaction, the excitations on these pairs are spectrally detuned from typical ions, see Fig.~\ref{fig:fig1}(a), which protects the pair excitations from decaying resonantly to their environment. For the thus protected coherent objects, a dominant source of dephasing originates from the energy shifts due to the virtual, circular exchange of an excitation between the pair ions and a neighboring ion ~\cite{PhysRevB.92.104428,dai2017four,PhysRevA.77.023603}. This ``ring-exchange" shift depends on the configuration of neighboring ions and thus senses their dynamics. Via its contribution to the dephasing of the pairs, the speed of the dynamics (the flipping rate) of the surrounding  $\text{Tb}^{3+}$ ions could thus be sensed.}

\section{Model and symmetries}
\label{sec:model}

To feature both, the interesting interaction-induced dynamics and highly coherent symmetry-protected cluster degrees of freedom, which may sense the former, a dipolar magnet requires certain general ingredients.

\subsection{Magnetically doped insulators}
\label{sec:1a}

For ions doped in a solid-state matrix, the crystalline environment breaks the rotational $SO(3)$ symmetry, and thus splits the angular momentum multiplets of a free ion into crystal-field (CF) levels. We focus here on rare-earth ions due to their compact $4f$ shell that is rather well-screened from interactions with other ions and lattice degrees of freedom. {For RE ions, the $\{L, S, J\}$ quantum numbers are quasi-conserved, while the continuous rotational symmetry associated with the conservation of $J_z$ is broken down to a set of discrete symmetries. If the ion has an odd number of electrons and thus $J$ is half-integer, the CF levels correspond to irreducible representations (irreps) of the local double group; as a consequence of Kramers' theorem, all levels are doubly degenerate. Such ions are referred to as ``Kramers ions". Conversely, non-Kramers ions have integer $J$ and an even number of $4f$ electrons, thus the CF levels correspond to irreps of the local point group.}

{
The dominant interactions among a dilute ensemble of magnetic ions are magnetic dipolar couplings. Those induce non-trivial quantum dynamics, provided they have matrix elements that are off-diagonal in the basis of the relevant low energy CF levels of the ions. Those indeed allow for resonant flip-flop processes (hopping of excitations between ions). Typically the temperature $T$ selects the CF states involved in the dynamics.
We call $\Delta$ and $\Delta_{\rm high}$, the first and second CF gaps, respectively. Provided the lowest two CF levels are coupled by the magnetic dipole operator, the condition $\Delta\lesssim T$ ensures interesting (equilibrium) dynamics.
If in addition $ T\ll\Delta_{\rm high}$ (as was realized in Ref.~\cite{beckert2024emergence}) we may restrict ourselves to these two lowest-lying levels. 
For simplicity, we will assume a CF level structure with  $\Delta \ll \Delta_{\rm high}$  henceforth.

{The coherent objects we target are symmetric clusters, whose symmetry protects the degeneracy of a doublet. For this symmetry not to be immediately spoiled by internal magnetic fields, we will see (cf.~Sec.~\ref{subsec:eff_int}) that it is essential that the relevant low energy CF levels be connected by matrix elements of a single component ($J_z$) of the total angular momentum only. As we will see, this ensures that a homogeneous field only couples like a scalar or pseudo-scalar to the cluster doublet, leaving its degeneracy intact. This condition rules out Kramers ions, where multiple components of $J_{x,y,z}$ have non-trivial matrix elements within or between CF doublets. %even though they generically will display interesting quantum dynamics.
We instead focus on materials containing non-Kramers ions as building blocks for the highly coherent clusters, while we do not rule out the presence of further Kramers ions, whose dynamics one might want to sense with clusters of non-Kramers ions.} The CF levels can either be singlets (one-dimensional irreps of the point group) or symmetry-protected multiplets. However, the latter usually carry magnetic moments, which couple to internal fields and are thus sensitive to magnetic noise. 
To protect from the associated dephasing,
we should require the relevant CF levels of the ions to be singlets. The angular momentum operator will then only act off-diagonally in the CF basis. 
The above discussion suggests that we restrict ourselves to insulators doped with non-Kramers RE ions featuring two point group singlets as ground state (GS) and first excited state (ES).

In order to allow for clusters of few ions of high symmetry, crystals of the tetragonal or hexagonal families are of special interest, since their point group symmetries allow for two-dimensional irreps, which enables the degeneracy of certain cluster states. At the same time, these symmetries ensure the condition that only the component of $\bm{J}$ along the principal axis, $J_z$, couples between the GS and ES singlets, while the matrix elements of $\{J_x,J_y\}$ vanish (as they belong to a doublet irrep of the point group)~\footnote{\label{footnote:rep}Consider two one-dimensional irreps $\Gamma_0$ and $\Gamma_1$, corresponding to CF singlets, and a two-dimensional irrep $\Gamma_2$, corresponding to the components $\{J_x,J_y\}$. The representation $\Gamma'=\Gamma_{1}\otimes\Gamma_2\otimes\Gamma_1$ cannot be reducible otherwise, upon tensoring with the adjoint of the one-dimensional $\Gamma_{GS}\otimes\Gamma_0$, one would obtain the contradiction that the irrep $\Gamma_d$ is reducible. It follows that $\Gamma'$ cannot contain the trivial representation, hence no non-zero matrix element can exist.}.

}

\subsection{Effect of the nuclear spin}
\label{sec:nucl_spin}

In nature, $3^+$ non-Kramers rare-earth ions always come with a half-integer, and thus non-zero, nuclear spin. The latter interacts with the electronic degrees of freedom via the hyperfine (HF) interaction. 
{In ions with the desired CF structure discussed above,} {this interaction hybridizes the lowest CF levels, giving rise to a finite magnetization (of order $\sim A_{\rm HF}/\Delta$) of the ion's electro-nuclear eigenstates, which increases the coupling to internal fields and magnetic noise.}
However, as the HF interaction couples the two levels only via $J_z$, its projection to the low-energy manifold is Ising-like, $\hat{H}_{\rm HF}=A_{\rm HF} J_z I_z$, where $I_z$ is the $z$-component of the nuclear spin and $A_{\rm HF}$ the HF coupling constant. $I_z$ is thus a quasi-conserved quantity, its value defining ``hyperfine species" of the non-Kramers ions.
A specific hyperfine species can be subjected to a so-called ``clock condition"~\cite{PhysRevLett.95.060502} by applying a longitudinal magnetic field $\bm{B}_{\rm ext}=-\bm{B}_{\rm HF}=-{A_{\rm HF} I_z}/(g_{J}\mu_B)$, where $g_{J}$ is the Landé $g$-factor, reinstating non-magnetic electro-nuclear wavefunctions on those ions.
Due to their different magnetizations, the various HF species interact differently with their environment, leading to different dynamics, {as we discuss in detail in App.~\ref{sec:supp_dyn_env}}. 

As far as protection from dephasing is concerned, one should always focus on ``clock-state ions" whose hyperfine field has been compensated. A single such ion is usually not yet a particularly good qubit as it may undergo resonant excitation flip-flops with neighboring ions. However, as we will show, substantially longer coherence times arise in small clusters of ions. {Furthermore, by probing the coherence of such clusters one may extract information on the dynamics of other components of the system, e.g., the flip rates of various HF species.}

\subsection{Cluster Hamiltonian}
\label{sec:H_cl}

Let us consider a cluster of $N_{\rm cl}$ ions located on spatially close sites of the regular host lattice, labeled by $i=1,...,N_{\rm cl} $. We denote the set of sites by $\Lambda_{\rm cl}$. We are particularly interested in clusters that have a non-trivial symmetry group $\mathcal{G}_{\rm cl}$, {which is a subgroup of the point group of its barycenter.} Note that in a randomly doped, dilute sample, with dopant concentration $\rho\ll 1$, the abundance of such clusters scales as $\sim\rho^{N_{\rm cl}}$, which is very small unless these clusters are implanted on purpose.~\footnote{With unpolarized nuclear spins, and if an additional requirement on the nuclear spin projections is relevant, the natural abundance of clusters further reduces to $\sim\left(\frac{\rho}{2I+1}\right)^{N_{\rm cl}}$.}

In general, the gap between crystal-field levels of doped ions depends slightly on local strains and other effects of nearby crystal defects. However, since we focus on compact symmetric clusters, we assume that the energy gap $\Delta_{\rm cl}= E_1-E_0$ between the {GS and ES singlets (with energies $E_0$ and $E_1$, respectively)} is the same for all cluster ions, and that the cluster symmetry is thus not lifted noticeably by such disorder effects. Note, however, that $\Delta_{\rm cl}$ in the cluster differs slightly from the average single ion splitting (which we called $\Delta$), because of the lattice distortions induced by the other cluster ions that substitute for the magnetically neutral RE ion of the host. The shifts $\Delta_{\rm cl}-\Delta$ are typically larger than the inhomogeneous broadening of single ions~\cite{zolnierek1984crystal,PhysRevLett.132.056703}. 

The ions in the cluster interact mostly via exchange and magnetic dipolar interactions. For every ion, we choose a local single-ion basis $\{\ket{0}_i,\ket{1}_i\}$ of the two lowest crystal-field singlets, along with associated triples of Pauli matrices $\bm{\sigma}_i$ acting in this low energy Hilbert space. 
To simplify our analysis, we choose the relative phase of the basis states such that the only non-zero matrix elements of the total angular momentum operator are {real and positive, $\mel{0}{J_z}{1}_i=\mel{0}{J_z}{1}>0$, independent of the ion.}
In this basis, we can write the interaction Hamiltonian as 
\begin{equation}
    \hat{H}_{\rm int}=\frac{1}{2}\sum_{i,j}J_{ij}\hat{\sigma}^x_i\hat{\sigma}^x_j,
\end{equation}
with the couplings
\begin{equation}
  J_{ij}=\frac{\mu_0(1-3\cos^2{\theta_{ij}})}{4\pi r_{ij}^3}m_{z}^2+\tilde{J}_{\text{ex},ij},
\end{equation}
where $\theta_{ij}$ is the angle between the distance vector $\bm{r}_{ij}$ and the $z$-axis, and $\tilde{J}_{\text{ex},ij}$ is the exchange contribution. We have also defined $m_{z}=g_J\mu_B\mel{0}{J_z}{1}$ as the matrix element of the magnetic moment along $z$.

Usually, the interactions are much weaker than the CF splitting, $|J_{ij}|\ll\Delta_{\rm cl}$, which justifies the use of the secular approximation, which drops terms that do not conserve the total number of excitations $M= N_{\rm cl}/2+S^z$, where $S^z= \frac{1}{2}\sum_i\hat{\sigma}_i^z$.
This approximation leads to the cluster Hamiltonian
\begin{equation}
  \hat{H}_{\rm cl}=\frac{1}{2}\sum_{i\in\Lambda_{\rm cl}}{\Delta_{\rm cl}\hat{\sigma}_i^z}+\frac{1}{2}\sum_{i,j\in\Lambda_{\rm cl}}J_{ij}(\hat{\sigma}_i^+\hat{\sigma}_j^-+\hat{\sigma}_i^-\hat{\sigma}_j^+),
  \label{eq:H_cl}
\end{equation}
{where $\sigma^{\pm}_i=\frac{\sigma^x_i\pm i\sigma^y_i}{2}$.}

The eigenstates of the cluster Hamiltonian $\hat{H}_{\rm cl}$ are characterized by the number of excitations $M$. The ground state is the only state in the $M=0$ manifold and corresponds to the product state $\ket{GS}_{\rm cl}=\bigotimes_{i=1}^{N_{\rm cl}}\ket{0}_i$. 
It is worth noting that, while the assumed conservation of $M$ simplifies our analysis, the symmetry considerations developed in the next section are valid for any $\mathcal{G}_{\rm cl}$-symmetric Hamiltonian.

\subsection{Cluster eigenstates}

\begin{figure}[ht!]
  \centering
  \includegraphics[width=0.48\textwidth]{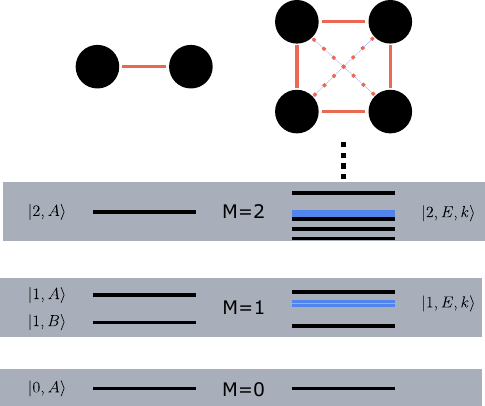}
  \caption{Energy level structure of two different ion clusters. On the left, a pair of ions with a $C_2$ cluster symmetry: all energy levels are singlets. On the right, the excitation manifolds $M=0,1,2$ of a quartet of ions with a $C_4$ symmetry: labeled and highlighted in blue are the symmetry-protected doublets ($k=1,2$). The irreps $\Gamma$ are denoted by Mulliken's notation.}
  \label{fig:fig1}
\end{figure}

The symmetry of the cluster implies that the Hamiltonian $\hat{H}_{\rm cl}$ commutes with the action of every element of its symmetry group $g\in \mathcal{G}_{\rm cl}$. This symmetry allows us to classify the eigenstates of $\hat{H}_{\rm cl}$ in terms of the irreps $\Gamma$ of $\mathcal{G}_{\rm cl}$.
Eigenstates belonging to the same irrep are degenerate. A given irrep may be realized multiple times on the cluster states, each having a distinct energy. 
We introduce an additional superscript $\alpha$ to distinguish such multiple occurrences for a given number $M$ of excitations. We thus label the eigenstates by $\ket{M,\Gamma^{(\alpha)},k}$, where $\Gamma^{(\alpha)}$ indicates the $\alpha$-th copy of irrep $\Gamma$ (using Mulliken's notation) in the $M$ manifold, while $k$ labels the degenerate states of irreps with dimension $>1$ (see Fig.~\ref{fig:fig1}). The eigenergies $\epsilon_{M,\Gamma^{(\alpha)}}$ do not depend on $k$. Any perturbation of the Hamiltonian that preserves the cluster symmetry cannot connect states belonging to different irreps. 

We are interested in the case where the cluster symmetry $\mathcal{G}_{\rm cl}$ admits 2-dimensional representations (doublets). Such degenerate states may be used to encode qubits or act as quantum sensors. In the next section, we show that these cluster states enjoy enhanced protection from various channels of decoherence, as compared to qubits formed merely from crystal-field levels of single ions or ion clusters with low symmetry, such as pairs.

\section{Coherence of qubits encoded in cluster doublets}
\label{sec:coherence}

{Here we focus on the manifold of cluster states with $M=1$ excitations {(unless otherwise stated)}, as they are simpler to prepare {in experiments} than those in manifolds of higher $M$, and also because the theoretical analysis simplifies. However, all results can  easily be extended to general $M$.}

Consider having prepared the wavefunction of a symmetric cluster as a specific quantum state within a cluster doublet $\Gamma_d$ belonging to the singly excited manifold $M=1$, with energy $\epsilon_d\equiv\epsilon_{1,\Gamma_d}$. We want to determine the coherence time of this qubit wavefunction (i.e., of the specific superposition of doublet basis states).
{Below, we develop a general theory of decoherence. After deriving the effective interaction between a cluster and {a neighboring single ion} (Sec.~\ref{subsec:eff_int}), we discuss how this coupling leads to both decay ($T_1$ effects) in Sec.~\ref{subsec:spectral_pr}, and pure dephasing ($T_2$ effects) in Secs.~\ref{subsec:ring_ex} and~\ref{subsec:dephasing}.} 
In Sec.~\ref{sec:even_cl}, a different suppression mechanism for clusters with an even number of ions is discussed.

\subsection{Effective interaction between a cluster doublet and its environment}
\label{subsec:eff_int}

We first derive the effective interaction between a cluster and a neighboring ion {(located on a site $l$)}.

We need to compute the transition matrix elements between different cluster eigenstates induced by interactions with neighboring ions. In order for the transformation under symmetry operations to become clear, we write this interaction explicitly in terms of the total angular momentum operators {
\begin{equation}
    \hat{H}_{l,\text{cl}}=\hat{J}^z_l\sum_{i\in\Lambda_{\rm cl}}D(\bm{r}_l - \bm{r}_i)\hat{J}^z_i,
    \label{eq:H_env-cl}
\end{equation}
}
where we denote by $\bm{r}_{l}$ the distance vector between the center of the cluster and the site $l$ and $D(\bm{r})=\frac{\mu_0 g_J^2}{4\pi}\frac{1-3\cos^2{\theta}}{|\bm{r}|^3}$, with $\theta$ the angle between the vector $\bm{r}$ and the $z$-axis. Note that this interaction only connects cluster states whose excitation numbers differ by $\pm 1$.

{Since the cluster diameter is usually small as compared to the distance $r_l$ to the neighbor ion $l$, we carry out a multipole expansion. Expanding the function $D(\bm{r})$ to the first order leads to the approximate Hamiltonian {
\begin{eqnarray}
    \hat{H}_{l,\text{cl}}&\approx &\hat{J}^z_l D({\bm{r}_{l}})\sum_{i\in\Lambda_{\rm cl}}\hat{J}^z_i\nonumber\\ 
  &&+\hat{J}^z_l\nabla D({\bm{r}_{l}})\cdot\sum_{i \in\Lambda_{\rm cl}}\bm{r}_{i}\hat{J}^z_i.
  \label{eq:multipole-exp}
\end{eqnarray}
}
}
The scaling of the effective coupling between ion $l$ and the cluster depends on whether or not these multipole terms have finite matrix elements between relevant cluster states.
To assess this, we have to analyze how the various terms of the multipole expansion transform under the action of a cluster symmetry $g\in \mathcal{G}_{\rm cl}$.

{The lowest (monopole) order transforms like the $z$ component of the total cluster angular momentum, whose one-dimensional irrep we call $\Gamma_{J_z}$.

For the dipole order, instead, we have to take into account the transformation of the cluster ions' coordinates. The coordinates transform according to the dipole charge representation $\Gamma_{\bm{r}}$. Accordingly the full dipole order transforms as $\Gamma_1=\Gamma_{J_z}\otimes\Gamma_{\bm{r}}$.

Higher ($n$-th) order multipole terms of $\hat{H}_{\rm env-cl}$ transform as the representation $\Gamma_{n}=\Gamma_{J_z}\otimes\Gamma_{\bm{r}^{(n)}}$, where $\Gamma_{r^{(n)}}$ is the representation of the $n$-th charge multipole \footnote{Note that, in the case of different ions (for instance Kramers ions) where also the matrix elements of the angular momentum components $J_x$ and $J_y$ between the considered CF levels are non-zero, additional interaction terms transforming as the representation $\Gamma_{\{J_x,J_y\}}\otimes\Gamma_{r^{(n)}}$ would be present. Such terms could lead to transitions between singlet and doublet cluster states even at the monopole order.}.}

\subsection{Spectral protection of clusters}
\label{subsec:spectral_pr}

The excitation energies on compact clusters of ions are detuned from those of isolated single ions
by an energy shift $\bar{J}$ of the order of the nearest-neighbor (dipole) interaction.
This detuning is typically much larger than the inhomogeneous broadening of the single ion spectrum (set by the disorder scale $W$) and the dipolar interaction $J_{\rm typ}$ with typical single ions close to the cluster, $\bar{J}\gg W, J_{\rm typ}$ {(at low concentration $\rho\ll 1$, the typical value of the dipolar interaction scales as $J_{\rm typ}\sim \rho\bar{J}$)}. As a consequence, cluster excitations cannot resonantly hop to neighboring ions. This spectral protection strongly suppresses the decay of the cluster excitation as compared to the decay rate $\kappa_{s}$ of similar excitations on single ions.

The finite decay rate $\kappa_s$ of excitations of neighboring ions induces a Lorentzian broadening of their spectral line, whose tail overlaps with the excitation energy of a cluster, where the Lorentzian has spectral weight $\sim\frac{\kappa_{s}}{\bar{J}^2}$. This opens a residual channel of decay, whose rate can be estimated via Fermi's Golden Rule. The flip-flop coupling between {an ion on site $l$ and the cluster transition $\ket{1,\Gamma_d,k}_{\rm cl}\to\ket{GS}_{\rm cl}$ is mediated by an effective interaction $J_{\rm eff}(\bm{r}_l)=\bra{GS}_{\rm cl}\bra{1}_l{\hat{H}_{l,\text{cl}}}\ket{0}_l\ket{1,\Gamma_{d},k}_{\rm cl}$} {(estimated below)}, where $\bm{r}$ is the distance between the ion and the center of the cluster. The decay rate induced by an ion in state $\ket{0}$ at a distance $\bm{r}$ is %with gap $\epsilon$ is
\begin{equation}
  \Gamma_{\rm cl}(\bm{r})\sim \frac{\abs{J_{\rm eff}(\bm{r})}^2}{\bar{J}^2}\kappa_{s},
\end{equation}
with $\kappa_{s}$ the single ion decay rate.

The effective interaction that mediates the de-excitation of the cluster from $M=1$ to $M=0$ is dominated by the leading non-zero matrix element of {the Hamiltonian of Eq.~\eqref{eq:H_env-cl}}, which arises from the smallest multipole order $n$ for which the tensor product $\Gamma_{GS}\otimes\Gamma_n\otimes\Gamma_d$ contains the trivial irrep, whereby $\Gamma_{GS}$ is the one-dimensional irrep carried by the cluster ground state. 
Now, the representation $\Gamma_{GS}\otimes\Gamma_{n=0}\otimes\Gamma_d$ cannot be reducible (for similar reasons as in footnote~tocounter{footnote}{-3}\footnotemark[\value{footnote}]tocounter{footnote}{+3}) and thus cannot contain the trivial representation. Therefore, the first non-vanishing matrix element arises only at a higher multipole order, $n\geq 1$.

The typical lifetime $T_1$ of cluster excitations is thus limited by the estimated decay rate
\begin{equation}
  T_{1,\rm cl}^{-1}=\Gamma_{\rm cl, typ}\sim \frac{\ev{J_{\rm eff}}_{\rm typ}^2}{\bar{J} ^2} \kappa_{s} \sim \left(\frac{J_{\rm typ}}{\bar{J}}\right)^2\left(\frac{a_{\rm cl}}{r_{\rm typ}}\right)^{2\nu} \kappa_{s},
\end{equation}
where $\nu$ is the index $n$ of the first term in the multipolar expansion that has a non-zero matrix element for the cluster transition that relaxes it to the ground state. The length $a_{\rm cl}$ is set by the cluster radius (of the order of the size of the unit cell), hence we expect the decay rate to decrease by a factor $\sim\rho^{2\nu/3}$ as compared to an unprotected excitation. For simple clusters, one usually finds $\nu=1$, as shown in App.~\ref{sec:supp_examples}. 

Note that, accordingly, the effective interaction of a cluster doublet with neighbor ions follows a power-law $J_{\rm eff}(\bm{r})\propto\partial^{(\nu)}_r f(\bm{r})\propto 1/r^{3+\nu}$, that falls off faster than the bare dipolar interaction.

\begin{figure}[ht!]
  \centering
  \includegraphics[width=0.48\textwidth]{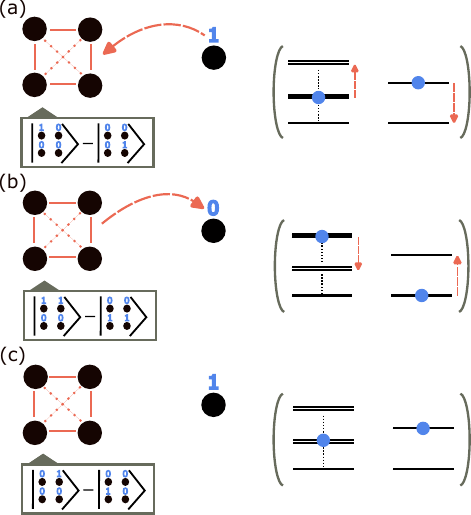}
  \caption{Schematic representation of one of the possible processes involved in ring-exchange interactions. Double lines in the spectra on the right indicate doublet states. On the left, the transfer of an excitation is indicated together with the initial state on the cluster, while on the right the transitions between the energy levels are shown. (a) A neighbor ion de-excites, while the cluster is excited to an intermediate state with $M=2$ or $M=0$ excitations. These virtual transitions are off-resonant by an amount proportional to the intra-cluster dipolar interactions. (b) The process inverse to (a), where, however, the cluster de-excites to a different state of the doublet. (c) The second order process where (a) is followed by (b) results in a final configuration with the same energy as the initial state, but the cluster will in general have changed the wavefunction within the doublet. The pictograms for cluster excitations used in this figure are explained in App.~\ref{sec:supp_examples}.}
  \label{fig:fig2}
\end{figure}

\subsection{Pure dephasing via ring-exchange}
\label{subsec:ring_ex}
Since we consider clusters of ions with non-magnetic eigenstates, magnetic noise does not dephase them to first order.
As was noticed in the study of dephasing of pair clusters of Tb$^{3+}$ in LiTb$_{x}$Y$_{1-x}$F$_4$~\cite{beckert2024emergence}, the leading contribution to pure dephasing of the clusters instead arises at second order, due to the virtual exchange of a cluster excitation with neighboring ions, see Fig.~\ref{fig:fig2}.
In the specific case of the symmetry-protected doublet $\Gamma_d$ in the excited manifold $M=1$, we obtain the exchange interaction with an isolated ion on a neighboring site $0$ by treating $\hat{H}_{\rm env-cl}$ perturbatively using a Schrieffer-Wolff transformation, as derived in App.~\ref{sec:supp_deriv},
{
\begin{eqnarray}
  \hat{V}_{\rm ex}({\bm{r}_{0}})=&&\begin{pmatrix}
  V_{d,-}({\bm{r}_{0}}) & V_{od,-}({\bm{r}_{0}}) \\
  V_{od,-}({\bm{r}_{0}}) & -V_{d,-}({\bm{r}_{0}}) \\
  \end{pmatrix}\otimes \hat{\sigma}^z_0\nonumber\\
  &&+\begin{pmatrix}
  V_{d,+}({\bm{r}_{0}}) & V_{od,+}({\bm{r}_{0}}) \\
  V_{od,+}({\bm{r}_{0}}) & -V_{d,+}({\bm{r}_{0}}) \\
  \end{pmatrix}\otimes \hat{\text{id}}_0\nonumber\\
  \equiv &&\hat{V}^{\rm dyn}({\bm{r}_{0}})\otimes\hat{\sigma}^z_0 + \hat{V}^{\rm st}({\bm{r}_{0}})\otimes\hat{\text{id}}_0,
  \label{eq:Vring}
\end{eqnarray}
with 
\begin{eqnarray}
  V_{d,\pm}({\bm{r}_{0}})=&&\frac{1}{4}\sum_{\psi}\left[\frac{\abs{\gamma_{1,\psi}^{(-)}}^2 - \abs{\gamma_{2,\psi}^{(-)}}^2}{\epsilon_d - \epsilon_\psi + \Delta_0 } \pm \frac{\abs{\gamma_{1,\psi}^{(+)}}^2 - \abs{\gamma_{2,\psi}^{(+)}}^2}{\epsilon_d - \epsilon_\psi - \Delta_0 }\right],\nonumber\\
  \label{eq:V_diag}\\
  V_{od,\pm}({\bm{r}_{0}})=&&\frac{1}{2}\sum_{\psi}\left[\frac{\gamma_{1,\psi}^{(-)}\left(\gamma_{2,\psi}^{(-)}\right)^*}{\epsilon_d - \epsilon_\psi + \Delta_0 }\pm\frac{\gamma_{1,\psi}^{(+)}\left(\gamma_{2,\psi}^{(+)}\right)^*}{\epsilon_d - \epsilon_\psi - \Delta_0 }\right],\nonumber\\
  \label{eq:V_offdiag}
\end{eqnarray}
where the sum runs over all cluster eigenstates $\ket{\psi} =\ket{M,\Gamma^{(\alpha)},k}$ that differ from the doublet, $\epsilon_\psi=\epsilon_{M,\Gamma^{(\alpha)}}$ being their energy. $\Delta_0$ denotes the CF splitting of the isolated ion ``$0$",
and the factors
\begin{equation}
  \gamma_{k,\psi}^{(\pm)}\equiv\sum\limits_{i\in\Lambda_{\rm cl}}J_{i,0}\mel{1,\Gamma_d,k}{\hat{\sigma}^\pm_i}{\psi}
  \label{eq:mat_el}
\end{equation}
are the transition matrix elements between the doublet and the states $\psi$.
These matrix elements can always be made real by {choosing doublet states with real coefficients in the chosen ion basis.}

Equation~\eqref{eq:Vring} shows that the interaction splits into two contributions: a ``dynamical" one, $\hat{V}^{\rm dyn}\otimes\hat{\sigma}^z_0$, which depends on the state of the external ion, and a  ``static" one $\hat{V}^{\rm st}\otimes\hat{\text{id}}_0$, which only depends on the position of the neighboring ion.
}
Since the effective exchange interaction Eq.~\eqref{eq:Vring} is not proportional to the identity operator on the cluster, it lifts the doublet degeneracy, as is to be expected because an additional dopant weakly breaks the symmetry that protected the cluster doublet. The presence of a nearby ion thus leads to the dephasing of the cluster doublet. It remains to estimate its strength.

We again expand the coupling $J(\bm{r})$ in Eq.~\eqref{eq:mat_el}, substitute it into Eqs.~(\ref{eq:V_diag},\ref{eq:V_offdiag}) and collect terms of equal order in the ratio $a_{\rm cl}/r$ to obtain an expansion of $V_{\rm ex}(\bm{r})$. Similarly as for decay processes, we find that the cluster symmetry suppresses the leading order term $\sim 1/r^6$. At long distances the ring-exchange decays with a power-law, $V_{\rm ex}(\bm{r})\propto 1/r^{6+\xi}$ where $\xi$ either takes the value $1$ or $2$, depending on selection rules. In the following, we classify the various situations that entail one or the other exponent. 

The value of $\xi$ depends on the order of the multipole expansion of $J(\bm{r})$ at which the cluster doublet is connected to other cluster states, and thus we need to establish how these terms transform under the cluster symmetry.
Using the monopole term for both excitation and de-excitation on the cluster preserves the degeneracy of the doublet and thus does not induce a splitting.

If instead for one of these steps the dipolar order ($n=1$) is used, the resulting interaction may split the doublet. {However, the splitting occurs only if the angular momentum component that connects the {doublet to the other cluster state}  transforms as a non-trivial irrep. If this is the case, then $\xi=1$.}
If instead at dipolar order there is either no coupling to a $M=2$ doublet that carries the same irrep $\Gamma_d$, or, alternatively, if the coupling is via a trivially transforming component of $J$, then both excitation and de-excitation steps require terms beyond the monopole order. 
{In such cases, the exponent $\xi=2$ arises. In App.~\ref{sec:supp_ring_exp}, a more formal treatment of the problem of determining the exponent $\xi$ is given.}
{Some representative cases of clusters and symmetry groups 
are discussed in detail in App.~\ref{sec:supp_examples}, where we find the exponent $\xi = 1$ in simple $C_3$- and $S_4$-symmetric clusters, while for a $C_4$-symmetric cluster we find $\xi=2$. We have not found larger exponents $\xi$ to occur in simple clusters.}

In conclusion, we find that for cluster states belonging to a doublet irrep, the ring-exchange is suppressed by a power-law decaying faster (with exponent $6+\xi$ with $\xi= 1,2$) than the spatial decay $\sim r^{-6}$ of ring-exchange interactions of qubits encoded in cluster states that are not symmetry-protected~\cite{beckert2024emergence}.

\subsection{Sensing of dynamical noise}
\label{subsec:dephasing}

As we have shown in the previous section, the ring-exchange interaction of a cluster with a neighboring ion contains both a static term and a dynamical term. Via the latter contribution, cluster states sense the dynamics of neighboring ions~\cite{beckert2024emergence}. By observing the decoherence of clusters, it is thus possible, in principle, to infer the flip rate $\kappa_s$ of typical ions, which in turn determines the energy diffusion in the dipolar system. {To actually determine $\kappa_s$, one needs a theory describing the coherence decay of cluster states. We derive this for an ensemble of symmetric clusters, surrounded by ions of density $\rho$.}  

We consider a cluster qubit under the perturbing effect of an ensemble of random fluctuators $\{s_{j}(t)\}_{j\notin\Lambda_{\rm cl}}$, which we assume to flip independently of each other with a flip rate $\kappa_{s}$. {For simplicity let us assume $T\gg \Delta$, such that the two thermally relevant CF levels are on average equally populated.} Each fluctuating neighbor ion thus produces telegraph noise on the cluster.

{Moreover, these ions break the symmetry of the cluster environment, which results in an additional static splitting of the doublet which is independent of any dynamics. This may mask the dynamical effect one is interested in. To mitigate such static (and other low frequency) sources of decoherence, one should use refocusing techniques, such as a CPMG sequence applied to the cluster~\cite{PhysRev.94.630,meiboom1958modified}. To implement this, the cluster qubit is subjected to a train of pulses, each implementing $\hat{\sigma}^y_{\rm cl}$} 
\footnote{Actually, as shown in Sec.~\ref{sec:state_manipulation}, four pulses are required to implement a single $\hat{\sigma}^y_{\rm cl}$. In this section by ``single pulse'' we refer to such a bundle of four sub-pulses. Indeed, conjugation with this operator inverts the sign of the interaction, $\hat{\sigma}^y_{\rm cl}V_{\rm ex}\hat{\sigma}^y_{\rm cl}=-V_{\rm ex}$, since we chose a basis in which $V_{\rm ex}$ is a real matrix. Thus its effect averages out over two pulse intervals.}

If the cluster is prepared in state $\ket{\psi_0}$ and subjected to a train of refocusing pulses, it decoheres into the state
\begin{equation}
  \ket{\psi(t)}=\hat{\mathcal{T}}\exp{-i\sum_{j\notin\Lambda_{\rm cl}}\int_0^t\hat{\phi}_j(t')\dd{t'}}\ket{\psi_0},
  \label{eq:psi_t}
\end{equation}
where $\hat{\mathcal{T}}$ is the time ordering operator and  
\begin{equation}
  \hat{\phi}_j(t)=\left(\hat{V}^{\rm dyn}(\bm{r}_j) s_j(t)+\hat{V}^{\rm st}(\bm{r}_j)\right)f(t)
  \label{eq:phi_op}
\end{equation}
is the phase contribution from the fluctuator at a distance $\bm{r}_j$, with the stepwise alternating function
\begin{equation}
  f(t)=\sum_{i=0}^{N_p}(-1)^i\left[\Theta(t-(2i-1)\tau) - \Theta(t-(2i+1)\tau)\right]
\end{equation}
that describes the reversal of the interaction sign, $\tau$ being the time delay between pulses {and $2N_p$ the total number of pulses.}

In Appendix~\ref{sec:supp_deph}, a detailed derivation of the coherence decay is presented, based on the calculation of the amplitude $\bra{\psi_0}\ket{\psi(t)}$, averaged over fluctuator positions and spin flip histories. One finds that the fidelity decays as
\begin{equation}
  \mathcal{F}(t)=\abs{\bra{\psi_0}\ket{\psi(t)}}^2\sim\exp{-\ev{V_\delta}_{\rm typ}^{3/\delta} G_{\delta}(t)},
  \label{eq:fid_decay}
\end{equation}
where $\delta = 6+\xi$ is the exponent governing the algebraic decay of ring-exchange (introduced in the previous section), $V_{\delta}(\bm{r})=\sqrt{\left[V_{d,-}(\bm{r})\right]^2 + \left[V_{od,-}(\bm{r})\right]^2}=V_0 g(\theta,\phi)/r^\delta$, where $\ev{V_{\delta}}_{\rm typ}\sim V_0\rho^{\delta/3}$ is the typical scale of the ring-exchange interaction and $G_{\delta}(t)=\ev{\abs{\int_0^t s(t')f(t')\dd{t'}}^{\frac{3}{\delta}}}_{s(t)}$ is a function of the spin flip statistics and the CPMG protocol. The notation $\ev{\dots}_{s(t)}$ indicates an average over all spin flip histories~\cite{bergli2009decoherence,PhysRevB.9.1,PhysRev.125.912}.

{We make the assumption (further discussed at the end of this section) that the interaction between a cluster and a neighboring ion is typically smaller than the flip rate $\kappa_s$. This assumption is well justified as long as the flip rate of typical ions is not heavily suppressed by strong disorder. Therefore, decoherence occurs on timescales $t\gtrsim 1/\ev{V_{\delta}}_{\rm typ}\gg 1/\kappa_s$, which is why we focus on the regime $\kappa_{s} t\gg 1$ below.} The short-time regime is described in App.~\ref{sec:supp_deph}. 

In the context of sensing, the flip dynamics of the neighboring ions is inferred from the decoherence of the cluster states. Considering a Hahn echo sequence ($N_p=1$), the fidelity decays as  
\begin{equation}
    \mathcal{F}(t)\sim\exp{-\left(\frac{t}{T_{\text{m.n.}}}\right)^{\frac{3}{2\delta}}},
    \label{eq:fid_T_mn}
  \end{equation}
with 
\begin{equation}
    T_{\text{m.n.}}^{-1}\sim \frac{\ev{V_{\delta}}_{\rm typ}^2}{\kappa_{s}}
\end{equation}
{(the subscript ``m.n." stands for ``motional-narrowing"~\cite{PhysRev.125.912})}. {The origin of the stretched exponential decay lies in the long coherence of rare clusters whose closest neighbors are atypically far away. Clusters with closest fluctuators at least a distance $r>r_{\rm typ}$ away, are subject to a total dephasing rate
\begin{equation}
    \gamma_d(r)\sim\rho\int_{r}^{\infty}\frac{V_\delta^2(r)}{\kappa_{s}}r^2\dd{r}\sim \frac{\ev{V_{\delta}}^2_{\rm typ}}{\kappa_{s}}\left(\frac{r}{r_{\rm typ}}\right)^{3 - 2\delta}
\end{equation}
and occur with a density $\sim e^{-\left(r/r_{\rm typ}\right)^3}$. Hence, at a time $t$, the ensemble fidelity is dominated by the clusters that optimize the product of the two exponentials, $e^{-\left(r/r_{\rm typ}\right)^3}e^{-\gamma_d(r)t}$, leading to the result in Eq.~\eqref{eq:fid_T_mn}.}
It applies unaltered also to the case of several pulses in the limit of long intervals between the pulses,  {$\kappa_{s}t\gg N_p$}, where a CPMG sequence is indeed ineffective.

Compared to an ``unprotected" qubit (with ring-exchange exponent $\delta =6$) there are two main differences that enhance the coherence: Since the stretching exponent depends inversely on $\delta$, a larger $\delta$ implies a slower echo decay. The increased ring-exchange exponent $\delta$ is also reflected in a stronger increase of the characteristic decoherence time scale with the inverse of the ion density $\rho$. Indeed, as compared to a symmetry-unprotected sensor, the characteristic decay rate is suppressed by an additional multiplicative factor $\sim\left(\frac{a_{\rm cl}}{r_{\rm typ}}\right)^{2\xi}\sim\rho^{2\xi/3}$.

To extend the coherence time beyond the value found above, one should utilize more frequent refocusing pulses, operating in the regime {$1\ll\kappa_{s} t\ll N_p$}. 
This leads to a slowed-down fidelity decay 
\begin{equation}
\label{eq:fid_T_N}
    \mathcal{F}(t)\sim\exp{-\left(\frac{t}{T_{N_p}}\right)^{\frac{9}{2\delta}}},
  \end{equation}
where the characteristic decoherence rate \begin{equation}
    {T}_{N_p}^{-1}\sim \left(\frac{\ev{V_{\delta}}_{\rm typ}^2 \kappa_{s}}{N_p^2}\right)^{1/3}
\end{equation} 
decreases with an increasing number of pulses $N_p$. Also in this case, this stretched exponential optimizes the product of two competing exponentials, one describing the rarity of clusters with atypically far neighbors, the other one capturing the slower temporal decay of the coherence of such clusters. In the regime of frequent refocusing, $\kappa_s\tau \ll 1$, the  phase accumulated over a time $t= N_p \tau$ from a single fluctuator at a distance $r$ scales only as $\abs{\delta\phi(t,r)} \sim V_{\delta}(r)\tau\sqrt{\kappa_{s} t}=V_{\delta}(r)\sqrt{\kappa_{s}}t^{3/2}/N_p$. {A number $\sim \rho r^3$ of such contributions add up randomly {(unlike before, the phase contributions from different neighbors are randomly signed), leading to a Gaussian suppression of the fidelity by a factor $e^{-\gamma_d(r)t}\sim\exp{- \rho r^3 \delta\phi^2(t,r)}$.}} As before, the competition between the decreasing decoherence with increasing $r$ from the nearest neighbors and the exponential rarity of finding clusters with no neighbors at shorter distances eventually leads to the result (\ref{eq:fid_T_N}).

{Let us define the coherence time $T_2$ as the time of e-fold decay of coherence. As we show in App.~\ref{sec:supp_deph}, this time scale either corresponds to $T_{\text{m.n.}}$ or $T_{N_p}$. We also discuss there that higher order corrections, arising from the non-commutativity of the ring-exchange interaction with different neighbors, might potentially alter the qualitative form of the long-time ($t\gg T_2$) decay of coherence, without, however, altering $T_2$.}

As mentioned earlier, the treatment of dephasing presented here assumes the weak coupling limit, $\ev{V_{\delta}}_{\rm typ}\ll \kappa_{s}$, in which it is safe to assume that the sensing cluster does not significantly affect the dynamics of its environment. In this limit the suppression of the ring-exchange interaction by the cluster symmetry enables one to probe much slower many-body dynamics {than would be possible without symmetry protection.}. In the opposite limit of strong coupling, one would need to determine whether the sensing cluster might instead act as an ``impurity potential". The latter might potentially freeze the coupled degrees of freedom by tuning them off-resonance with respect to the bulk of the system, thus altering the intrinsic dynamics that the cluster is intended to sense.

\subsection{ Extra protection from ``same species" neighbors in clusters with an even number of ions}
\label{sec:even_cl}

There is an additional mechanism that protects certain cluster states from dephasing by the fluctuations of the same HF species of ions as constitute the cluster. This mechanism is independent of cluster symmetry but only occurs in clusters with an even number of ions. 

The ring-exchange coupling is found to be suppressed for cluster eigenstates in the manifold with $M=N_{\rm cl}/2$ excitations, whereby this suppression only occurs for exchange with neighbor ions that share the same CF splitting as the cluster ions, which are assumed to be in a clock-state condition. In the presence of different hyperfine species, this extra protection increases the coherence significantly only if either the HF interaction is negligible, or if only the clock-state ions have any significant dynamics.   

The above result relies on the secular approximation and an ensuing effective particle-hole symmetry in the $M=N_{\rm cl}/2$ manifold. Indeed, in that manifold the first term of $\hat{H}_{\rm cl}$ (cf.~Eq.~\eqref{eq:H_cl}) is zero, while the second term, which we call $\hat{H}_{\rm int}$, commutes with the operator $X=\prod_{i\in\Lambda_{\rm cl}}\hat{\sigma}_i^x$, which inverts the excitation state of all ions.
This implies that cluster eigenstates in this manifold are either odd or even under $X$. Let us now consider the second-order energy shift of a cluster eigenstate $\ket{\alpha}\equiv\ket{N_{\rm cl}/2,\Gamma}$ due to a ring-exchange with a neighbor ion on a site $0$. The latter has gap $\Delta_0$ and can be either in state $\ket{0}$ or $\ket{1}$. The associated energy shifts which they induce will be labeled by superscripts $\mp$, respectively. They are easily obtained in second order of perturbation theory as 
\begin{eqnarray}
    &&\delta E_{\alpha}^{\pm}=\ev{\sum_{i\in\Lambda_{\rm cl}}J_{i,0}\hat{\sigma}_i^{\mp}\frac{1}{E_{\alpha}\pm \Delta_0 -\hat{H}_{\rm cl}}\sum_{i\in\Lambda_{\rm cl}}J_{i,0}\hat{\sigma}_i^{\pm}}{\alpha}\nonumber\\
    &&=\ev{\hat{\Sigma}^{\mp}\frac{1}{E_{\alpha}\pm \Delta_0 \mp \Delta_{\rm cl} -\hat{H}_{\rm int}}\hat{\Sigma}^{\pm}}{\alpha},
\end{eqnarray}
where we have defined the operator $\hat{\Sigma}^{\pm}\equiv\sum_{i\in\Lambda_{\rm cl}}J_{i,0}\hat{\sigma}_i^{\pm}$.
Exploiting the aforementioned symmetry with respect to $X$ (and using the fact that it squares to the identity), we obtain
\begin{eqnarray}
    \delta E_{\alpha}^+&&=\ev{\Sigma^{-}\frac{1}{E_{\alpha}+ \Delta_0 -\hat{H}_{\rm cl}}\Sigma^{+}}{\alpha}\nonumber\\
    &&=\ev{X\Sigma^{-}\frac{1}{E_{\alpha}+ \Delta_0 -\hat{H}_{\rm cl}}\Sigma^{+}X}{\alpha}\nonumber\\
    &&=\ev{X\Sigma^{-}X\frac{1}{E_{\alpha} + \Delta_0 -\Delta_{\rm cl}- X\hat{H}_{\rm int}X}X\Sigma^{+}X}{\alpha}\nonumber\\
    &&=\ev{\Sigma^{+}\frac{1}{E_{\alpha} - \hat{H}_{\rm int} - \Delta_{\rm cl} +\Delta_0}\Sigma^{-}}{\alpha}\nonumber\\
    &&\approx\delta E^{-}_\alpha +\ev{\Sigma^{+}\frac{\Delta_{\rm cl}-\Delta_0}{\left(E_{\alpha}-\hat{H}_{\rm int}\right)^2}\Sigma^{-}}{\alpha},
\end{eqnarray}
where we have neglected terms $\sim O\left(\left(\frac{\Delta_{\rm cl}-\Delta_0}{\bar{J}}\right)^2\right)$, {$\bar{J}$ being the scale of nearest-neighbor dipole interactions, as introduced in Sec.~\ref{subsec:spectral_pr}}.
We see that for identical CF splittings on the cluster and neighboring single ions, $\Delta_{\rm cl}=\Delta_0$, the second order shifts are identical, $\delta E^+=\delta E^- $, and thus do not depend on the state of the neighboring ion. A qubit encoded in the superposition of two different cluster singlets within the $M=N_{\rm cl}/2$ manifold {thus couples to neighboring ions that have the same CF gap as the cluster ions} only via higher order processes or via corrections to the secular approximation which are of order $\propto {\bar{J}}/{\Delta_{\rm cl}}$. This implies in turn that such mid-spectrum states couple more strongly to ions with a different CF gap, which may be interesting in the context of quantum sensing of the dynamics of those species. 

{We caution, however, that the CF splittings $\Delta_{\rm cl}$ typically differ from $\Delta_0$ by an amount $\sim \delta Q\Delta_0$}, due to level shifts caused by the presence of the other cluster ions, as we discussed in Sec.~\ref{subsec:CF_deform}. This implies that the ring-exchange dephasing due to like HF-species is not fully eliminated, but merely reduced by an extra factor $\propto\delta Q\frac{\Delta_0}{\bar{J}}$, which is not necessarily very small.

In the case of doublets, an analogous analysis shows that only the static part $\hat{V}^{\rm st}(\bm{r})$ of the interaction with a neighboring ion survives if $\Delta_{\rm cl}=\Delta_0$, which can thus be eliminated by a CPMG sequence.

\section{State preparation and manipulation}
\label{sec:driving}
To carry out quantum sensing it is essential to be able to prepare and manipulate specific quantum states in multiplets such as the doublet $\Gamma_d$ considered above.
What makes this challenging is that the cluster ground state is a singlet, and hence, to drive a transition to a doublet one needs to act with an operator that transforms in a doublet representation under $\mathcal{G}_{\rm cl}$. Natural candidates of such operators are the $x,y$-components of the magnetic or electric dipole moment. {However, when acting on the ground state singlet of an isolated ion these operators cannot couple it to the first excited singlet we have been considering \footnote{We note that the $x,y$-components of the electric dipole also constitute a doublet irrep.}. 

At first sight, it would thus seem impossible to couple the GS to cluster doublets in higher $M$-manifolds. However,} there is a way out for ions in a cluster, where the other cluster ions weakly deform the crystal-field potential, thereby lowering the symmetry and lifting the selection rules for the single ion and thus allowing for such matrix elements. This enables a finite, albeit weak, coupling between the cluster GS and a cluster doublet in the $M=1$ manifold.
This can be done, e.g., using the magnetic field of a single resonant microwave pulse to drive a cluster from its ground state to a specific state in the doublet $\Gamma_d$ in the manifold with $M=1$ excitation. An alternative route passes through {highly excited states of opposite parity (e.g. $4f^{n} \to 4f^{n-1}5d$) as intermediate steps}. Since these are electric rather than magnetic dipole transitions, they can be substantially faster. The drawback is that the intermediate states have a short lifetime which limits the fidelity of this state preparation. 

Below we discuss these two routes in some more detail.

\subsection{Deformation of the CF potential due to the presence of other cluster ions}
\label{subsec:CF_deform}
A dopant ion deforms the crystal and changes the charge distribution in its vicinity. This induces a perturbation to the CF Hamiltonian of nearby magnetic ions~\cite{PhysRevLett.132.056703,zolnierek1984crystal} as compared to the Hamiltonian of a single isolated dopant. The perturbing potential can be approximately described by a point charge model, assuming an effective excess charge $\delta Q$ on the sites of nearby dopant ions. 
To obtain an order of magnitude estimate, we may assume a typical order of magnitude of $\delta Q \sim 1\%-10\%$~\cite{zolnierek1984crystal} of an electron charge.

Accordingly, the point symmetry of a selected ion in a cluster is weakly broken due to the potentials induced by the other ions. These perturbations allow for non-zero matrix elements (of order $O(\delta Q)$) for otherwise prohibited single ion transitions, especially between two CF singlets of the given ion $\mel{1}{J_{x/y}}{0}_i = O(\delta Q)$.

In the following, we consider such weakly allowed matrix elements for the components $\{J_x,J_y\}$ of the total angular momentum between the two lowest CF singlets of the ions, and later for in-plane electric dipole moments. 

\subsection{Cluster state preparation via a single pulse}

We assume a generic pulse propagating along the $z$-axis, whose magnetic field drives a magnetic dipole transition between the two considered CF singlets, coupling to the ions via the operator 
\begin{equation}
    \mathcal{M}=\Omega(t) \left[h_x J_x + h_y J_y\right] e^{i\omega t} + h.c.,
\end{equation}
where $\Omega(t)$ is a generic envelope, $\omega=\epsilon_d-\epsilon_{GS}$ and $\{h_x,h_y\}$ are the magnetic field components. When coupling to the cluster, we obtain a time-dependent contribution to the cluster Hamiltonian containing terms $\hat{H}_{\rm int}(t)=\sum_{i\in\Lambda_{\rm cl}}\{\Omega(t)\left[h_x \mel{1}{J_x}{0}_i + h_y \mel{1}{J_y}{0}_i\right] e^{i\omega t}\hat{\sigma}^+_i+\dots\}$, from which one sees that the driving term carries the same representation as the components $\{J_x,J_y\}$.
{If we write the decomposition of the two $M=1$ doublet states as
\begin{equation}
    \ket{1,\Gamma_d,k}=\sum_{i\in\Lambda_{\rm cl}}\left(\psi_{k}(i)\ket{1}_i\bigotimes_{j\neq i}\ket{0}_j\right),
\end{equation}
we can write the effective coupling between the cluster ground state and the doublet as
\begin{eqnarray}
  \mathcal{M}_{\rm eff}=\left(\mathcal{A}_{\bm{h}}\ket{1,\Gamma_d,1}+\mathcal{B}_{\bm{h}}\ket{1,\Gamma_d,2}\right)\bra{GS} + h.c.,\nonumber\\
  \label{eq:Meff}
\end{eqnarray}
where
\begin{eqnarray}
  \mathcal{A}_{\bm{h}}&&=\Omega(t)e^{-i\omega t}\sum_{i\in\Lambda_{\rm cl}}\left[h_x \mel{1}{J_x}{0}_i + h_y \mel{1}{J_y}{0}_i\right]\psi^{*}_{1}(i),\nonumber\\
  \mathcal{B}_{\bm{h}}&&=\Omega(t)e^{-i\omega t}\sum_{i\in\Lambda_{\rm cl}}\left[h_x \mel{1}{J_x}{0}_i + h_y \mel{1}{J_y}{0}_i\right]\psi^{*}_{2}(i).\nonumber\\
\end{eqnarray}
}

By choosing the amplitudes of $h_x$ and $h_y$ and their relative phase, and adjusting the length of the pulse to a $\pi$-pulse, arbitrary doublet states can be prepared. For a pulse of a given intensity, the Rabi frequency of the nearly forbidden transition is reduced by a factor $\sim\delta Q$ as compared to typical single ion transitions that are magnetic-dipole allowed.

\subsection{Two pulses}
The reduction $\propto\delta Q$ of the Rabi frequency is unavoidable. However, one may try to compensate for its weakness by using a two-pulse drive via higher lying states, the transition to which are electric dipole allowed, and thus much stronger than the magnetic dipole transitions considered above. Specifically, let us consider the manifold of $4f^{n-1}5d$ states of single ions. Those have parity opposite to the low energy $4f^n$ states, and hence $4f^{n}\to 4f^{n-1}5d$ transitions are electric dipole allowed. Those are typically two orders of magnitude stronger than magnetic dipole transitions. Within the excited manifold, it is preferable to consider the lowest single ion CF state $\ket{e}$ (having energy $\epsilon_e$), because of its longer lifetime. For simplicity, we assume it to be a singlet.

A two-pulse driving strategy will first excite the cluster into the state $\sum_{i\in\Lambda_{\rm cl}}\left(\ket{e}_i\bigotimes_{j\neq i}\ket{0}_j\right)$, {of energy $\epsilon_{e,\text{first}}$}, by driving with an electric field along $z$, with a frequency resonant with the transition $\omega_1=\epsilon_{e,\text{first}}-\epsilon_{GS}$.
We may assume that this singlet-to-singlet transition is allowed without the need for weak CF symmetry breaking. {Driving with a Rabi frequency well below the dipolar splitting of cluster levels with fixed $M$ will ensure that the excitation to higher manifolds $M>1$ remains off-resonant. }Subsequently, a second pulse polarized in the $xy$-plane drives this excited state into a specific superposition of the $\Gamma_d$ doublet states.
The pulse frequency must now be resonant with this different transition, $\omega_2=\epsilon_{e,\text{first}} - \epsilon_d$.
Just as before, the choice of polarization allows one to prepare a generic superposition of the doublet states. In this case, however, the effective Rabi frequency, while still reduced by a factor $\delta Q$, is substantially faster because of the larger matrix element of the electric dipole transition.

\subsection{State manipulation}
\label{sec:state_manipulation}
To carry out refocusing techniques (cf.~Sec.~\ref{subsec:dephasing}) it is important to be able to perform arbitrary 1-qubit gates, such as  $\sigma_y$, on a cluster doublet state. 

We assume again that the cluster doublet $\Gamma_d$  belongs to the manifold with $M=1$ excitation.
An arbitrary gate can be implemented with four pulses with magnetic fields in the $xy$-plane. A first $\pi$-pulse transfers a selected doublet state into a singlet in the $M=2$ manifold. The initial state is selected by the choice of polarization, similar to what was shown for the excitation from the ground state to a doublet in Eq.~\eqref{eq:Meff}. In the same fashion, a second $\pi$-pulse drives the other doublet state to a different singlet. Finally, with two further $\pi$-pulses, the states in these two singlets can be transferred back to the doublet $\Gamma_d$, with the possibility of exchanging them and introducing any desired phase by tuning the polarization, duration, phase and delay between the pulses.

{The duration of any such gate operation is bounded from below by the inverse of the Rabi frequency of these pulses. A priori, a strong enough drive may implement the proposed refocusing technique fast enough (ideally, faster than the flip rate $\kappa_s$, {so as to suppress environmental decoherence}) and with high fidelity.}

\section{Conclusion} 
\label{sec:conclusions}
{In this paper we have considered the problem of quantum coherence and dynamics in magnetically doped insulators. In general the mutual interaction between the close ions of a cluster leads to spectral detuning, which suppresses resonant decay to typical, more isolated ions and thus prolongs the coherence time of cluster states.
We have shown that, in addition, in crystals doped with non-Kramers RE ions, under specific conditions on their CF spectrum, quantum coherence can be significantly enhanced in high-symmetry clusters of ions, compared to single ions or clusters of ions of low or no symmetry (such as pairs).
Their coupling to other ions is significantly reduced, which enhances both the $T_1$ and $T_2$ times of a qubit built from those multiplets. 

We have shown how quantum states in cluster doublets can be prepared and controlled, which is fundamental for both quantum memory and quantum sensing applications. 
However, increased protection from environmental noise inevitably comes at the price of reduced coupling to external driving fields used for quantum control. We have shown, though, that coherent preparation and manipulation of the protected quantum states is nonetheless possible using various pulse driving schemes.

Concerning the use of such clusters for quantum sensing, we have shown that the dephasing of properly prepared quantum states of symmetric clusters can sense the speed of the non-trivial many-body dynamics of the surrounding disordered dipolar magnet.

Finally, we recall that in the present work we have considered randomly doped materials. In those, any type of cluster will occur, albeit with a low abundance $\sim\rho^{N_{\rm cl}}$ that decreases exponentially with the cluster size. This makes it very challenging to detect the signal from their (generalized) Hahn echoes above the noise level. 
However, future technologies~\cite{diller2019magnetic,dreiser2015molecular,affronte2007single} might allow the implantation of such clusters deterministically on surfaces, so that their density could be much higher.}

{\section{Acknowledgments}
We acknowledge financial support from the Swiss National Science Foundation under Grants Nos.~166271 and 200558. We thank Gabriel Aeppli for discussions.
}
\appendix

\section{Dynamics in the environment of a cluster}
\label{sec:supp_dyn_env}
The considered cluster is embedded in an insulating crystal. A typical fabrication process will result in a dilute set (of density $\rho$) of single ions of the same type in the surroundings of the cluster. One leading source of decoherence for quantum states of the cluster is due to their interaction with these other ions. Phonon-induced relaxation is generally negligible for a small gap $\Delta$ (in the microwave range), as the wavelength of phonons of energy $\sim \Delta$ is much larger than the radius of the ionic wavefunctions and the density of phonon states is low.

The $N$ dopant ions in the crystal {(including all hyperfine species) are described the following Hamiltonian
\begin{equation}
  \hat{H}_{\rm full}=\sum_{n=1}^{N}{\left[\frac{\Delta_n}{2}\hat{\sigma}_n^z + h_n(I_z)\hat{\sigma}_n^x\right]}+\frac{1}{2}\sum_{n,m}J_{nm}\hat{\sigma}_n^x\hat{\sigma}_m^x,
  \label{eq:H_full}
\end{equation}
where $h_n = ({A_{\rm HF}I_z} + g_J\mu_B B_z)\mel{0}{J_z}{1} +\delta h_n$ accounts for the HF interaction, the Zeeman coupling to the external field $B_z$ and additional internal fields ($\delta h_n$) originating from further electronic or nuclear spins in the host crystal that were not yet accounted for in the Hamiltonian.}

Due to strain and crystal defects the ions' CF splittings $\Delta_n=\Delta + \delta\Delta_n$ have a small random component {$\delta\Delta_n$, which we assume to be independently distributed with variance $W^2_{\rm CF}$, characterizing the CF disorder.} 

As mentioned in Sec.~\ref{sec:nucl_spin}, different hyperfine species are subject to different dynamics. This can be seen if we rotate the basis of each ion to diagonalize the single-ion terms in Eq.~\eqref{eq:H_full}. In this new basis, which we denote by $\{\ket{\uparrow}_n,\ket{\downarrow}_n\}$, the total angular momentum $J_z$ has both diagonal and off-diagonal matrix elements,
\begin{eqnarray}
    \abs{\mel{\uparrow}{J_z}{\uparrow}_n}&&=\frac{\abs{h_n(I_z)}}{\sqrt{\Delta_n^2 + h_n(I_z)^2}}\equiv m^{(d)}_{n}(I_z),\\
    \abs{\mel{\uparrow}{J_z}{\downarrow}_n}&&=\frac{{\Delta_n}}{\sqrt{\Delta_n^2 + h_n(I_z)^2}}\equiv m^{(od)}_{n}(I_z).
\end{eqnarray}
We  can thus rewrite the Hamiltonian as
\begin{eqnarray}
    \hat{H}_{\rm full}=&& \frac{1}{2}\sum_{n=1}^{N}{\tilde{\Delta}_n\hat{\tau}_n^z}+\frac{1}{2}\sum_{n,m}\left(J^{xx}_{nm}\hat{\tau}_n^x\hat{\tau}_m^x + J^{zz}_{nm}\hat{\tau}_n^z\hat{\tau}_m^z\right)\nonumber\\
    &&+\frac{1}{2}\sum_{n,m}\left(J^{xz}_{nm}\hat{\tau}_n^x\hat{\tau}_m^z + J^{zx}_{nm}\hat{\tau}_n^z\hat{\tau}_m^x\right),
\end{eqnarray}
where $\bm{\tau}_n$ are Pauli matrices in the new basis, $\tilde{\Delta}_n=\sqrt{\Delta_n^2 + h_n(I_z)^2}$, $J^{xx}_{nm}=J_{nm}m^{(d)}_{n}m^{(d)}_{m}$, $J^{zz}_{nm}=J_{nm}m^{(od)}_{n}m^{(od)}_{m}$ and $J^{xz}_{nm}=J_{nm}m^{(d)}_{n}m^{(od)}_{m}$. 

Non-clock ions may carry a non-negligible magnetization. As a consequence, the total disorder seen by these ions is larger, owing to their coupling to internal fields, $W^2_{\text{non-clock}, I_z}\sim W^2_{\rm CF} + \left(m_n^{(d)}(I_z)\right)^2\ev{\delta h_{\text{tot}, n}^2}$. {The internal field $\delta h_{\text{tot}, n}$ contains contributions from the dipolar field of neighboring dopants (of order $\sim J_{\rm typ}$, cf.~Sec.~\ref{subsec:spectral_pr}) and from other sources of magnetic noise in the system.} As a result of the increased disorder, excitations on these ions tend to be more localized than clock ion excitations. {However, since these internal fields are dynamical, non-clock ions decohere much faster than clock ions.} 

In contrast, for clock-state ions, $h_n(I_z)=\delta h_n$ is usually negligible (as compared to $\Delta$), which results in a much smaller residual magnetization as compared to non-clock ions. As a result typical clock-state ions decohere predominantly due to excitation hopping to {other clock-state ions}.
Furthermore, the internal fields which mostly originate from other dopants, do not contribute significantly to the total disorder of clock-state ions, $W^2_{\rm clock} \approx W^2_{\rm CF}$.

\section{Derivation of ring-exchange interaction of a cluster with a neighbor ion}\label{sec:supp_deriv}
We consider a system consisting of a symmetric cluster and one additional neighboring ion. We define the subspace $\mathcal{P}$, spanned by the cluster doublet $\Gamma_d$ (of energy $\epsilon_{1,\Gamma_d}\equiv\epsilon_d$) tensored with the Hilbert space of the additional ion, $\{\ket{1,\Gamma_0,k}\otimes\ket{n}\equiv\ket{k,n}\}_{k=1,2;n=0,1}$. The total Hilbert space $\mathcal{H}$ is the direct sum of $\mathcal{P}$ and its orthogonal complement $\mathcal{P}^{\perp}$.

We now carry out a Schrieffer-Wolff transformation with respect to this separation.
By truncating the perturbative expansion at the second order~\cite{bravyi2011schrieffer}, we obtain the effective interaction Hamiltonian in the subspace $\mathcal{P}$
\begin{eqnarray}
  \mel{k,n}{\hat{H}_{\rm eff}}{k',n'}=&&\sum_{j}\bra{k,n}{\hat{H}_{\rm env-cl}}\dyad{j}{j}{\hat{H}_{\rm env-cl}}\ket{k',n'}\nonumber\\
  &&\times\frac{1}{2}\left(\frac{1}{E_{k,n} - E_j} + \frac{1}{E_{k',n'} - E_j}\right),
  \label{eq:app_H_eff}
\end{eqnarray}
where the sum runs over the eigenstates $j\in \mathcal{P}^{\perp}$ of the decoupled cluster-ion system. Here the unperturbed energies in ${\cal P}$ are $E_{k,n}=\epsilon_d - (-1)^n\frac{\Delta_0}{2}$, where $\Delta_0$ is the CF gap of the additional ion.

We now write the $4\times 4$ matrix $\hat{H}_{\rm eff}$ as a sum over tensor products $\hat{\sigma}_{\rm cl}^\alpha\otimes\hat{\sigma}_0^\beta$ of $2\times 2$ matrices that act on the cluster doublet and the single ion, respectively. {We use $\hat{H}_{\rm env-cl}$ from Eq.~\eqref{eq:H_env-cl}, assuming the validity of the secular approximation. This implies} that $\mel{k,n}{\hat{H}_{\rm eff}}{k',n'}=0$ if $n\neq n'$, and thus only operators $\propto \hat{\sigma}_0^z$ or $\propto \hat{\text{id}}_0$ that preserve the excitation state of ion $0$ occur. 
{From this we obtain}
\begin{eqnarray}
  \hat{V}_{\rm ex}(\bm{r})=\left[V_{d,-}(\bm{r})\hat{\sigma}_{\rm cl}^z + V_{od,-}(\bm{r})\hat{\sigma}_{\rm cl}^x\right]\otimes\hat{\sigma}_0^z \nonumber\\
  +\left[V_{d,+}(\bm{r})\hat{\sigma}_{\rm cl}^z + V_{od,+}(\bm{r})\hat{\sigma}_{\rm cl}^x\right]\otimes \hat{\text{id}}_0^z,
\end{eqnarray}
where the coupling functions $V_{d,\pm}(\bm{r})$ and $V_{od,\pm}(\bm{r})$ have been given in Eqs.~(\ref{eq:V_diag},\ref{eq:V_offdiag}) and we have dropped terms that act as the identity on the cluster doublet and thus do not dephase it. This corresponds to Eq.~\eqref{eq:Vring}. The subscripts $d,od$ refer to the terms being diagonal or off-diagonal with respect to the cluster doublet basis, respectively, while
the label $\pm$ indicates whether the interaction depends on the excitation state of the neighbor ion (-) or not (+).

Note that terms acting as $\propto \hat{\sigma}_{\rm cl}^y$ on the cluster do not appear if for the doublet $\Gamma_d$ one chooses a basis with real coefficients in the ions' product state basis.

\section{Dephasing by random fluctuators}\label{sec:supp_deph}

In this section, we derive the decay of the amplitude decay given in Eq.~\eqref{eq:fid_decay} due to the dephasing effect of random fluctuators.
 {The behavior of the complex time-ordered exponential in Eq.~\eqref{eq:psi_t} can be relatively easily understood at a qualitative level if we disregard the commutators of the ring-exchange interactions pertaining to different ions. We will do so below. At the end of this appendix, we will show that under certain conditions the neglected terms do not significantly alter the decay.} We then approximate the amplitude $\bra{\psi_0}\ket{\psi(t)}$ (cf.~Eq.~\eqref{eq:psi_t}) as
\begin{equation}
  \bra{\psi_0}\ket{\psi(t)}\approx\ev{\prod_{j}\exp{-i\int_0^t\hat{\phi}_j(t')\dd{t'}}}{\psi_0}.
  \label{eq:ampl_approx}
\end{equation}

{Recall that the operators $\hat{\phi}_i$, defined in Eq.~\eqref{eq:phi_op}, are represented by $2\times 2$ matrices in the space of the cluster doublet.
We now take an ensemble average over all fluctuator histories and positions, since we seek to describe the dephasing averaged over many otherwise equivalent clusters. As we are considering a CPMG sequence of pulses, the total time $t$ is an integer multiple of $\tau$. The  average over ion positions is done by subdividing the volume into infinitesimal volume elements $\dd{V}_j$, which contribute with an exponential $\exp{-i\int_0^t\hat{\phi}_j(t')\dd{t'}}$ with probability $\rho\dd{V}_j$ (i.e., only when occupied by an ion), then taking the continuum limit~\cite{grimm2023quantum}}
\begin{equation}
  \bra{\psi_0}\ket{\psi(t)}\approx\ev{\exp{\rho\int\dd^3{r}\left(\ev{e^{i\hat{\Phi}(\bm{r},t)}}_{s(t)}-1\right)}}{\psi_0},
\end{equation}
where
\begin{equation}
    \hat{\Phi}(\bm{r},t)=-\int_0^t{\hat{\phi}(\bm{r},t')\dd{t'}}=-\hat{V}^{\rm dyn}(\bm{r})\int_0^t s(t')f(t')\dd{t'}.
\end{equation}
Note that by construction of the echo sequence, the static term proportional to $\hat{V}^{\rm st}(\bm{r})$ averages to zero, since $\int_0^t f(t')\dd{t'}=0$ for $t$ being integer multiples of $\tau$.

We now write 
\begin{equation}
  e^{i\hat{\Phi}({\bm{r}},t)}=\cos{(h(\bm{r},t))}\hat{\text{id}}_0 +i \sin{(h(\bm{r},t))}\bm{n}(\bm{r},t)\cdot\hat{\bm{\sigma}},
  \label{eq:exp_decomp}
\end{equation}
with $\hat{\text{id}}_0$ the $2\times 2$ identity matrix in the doublet space, and
\begin{eqnarray}
  h(\bm{r},t)=&&\sqrt{\left[V_{d,-}(\bm{r})\right]^2 + \left[V_{od,-}(\bm{r})\right]^2}\int_0^t s(t')f(t')\dd{t'}\nonumber\\
  =&&V_0\frac{g(\theta,\phi)}{r^{\delta}}\int_0^t s(t')f(t')\dd{t'},
\end{eqnarray}
where $g(\theta,\phi)$ is a non-negative adimensional function containing the angular dependence of the coupling. The Bloch sphere unit vector is $\bm{n}(\bm{r},t)=\frac{-1}{\sqrt{\left[V_{d,-}(\bm{r})\right]^2 + \left[V_{od,-}(\bm{r})\right]^2}}\{V_{d,-}(\bm{r}),0,V_{od,-}(\bm{r})\}$.

When taking the average over spin flip histories of Eq.~\eqref{eq:exp_decomp}, it is easily seen that only the first term survives, as it is even under spin inversion ($s(t)\to -s(t)$). We are left with
\begin{equation}
  \bra{\psi_0}\ket{\psi(t)}\approx\exp{\rho\int\dd^3{r}\left(\ev{\cos{h(\bm{r},t)}}_{s(t)}-1\right)}.
\end{equation}
Finally, we perform the spatial integration and take the absolute value squared to obtain the fidelity
\begin{eqnarray}
  &&\mathcal{F}(t)= \bra{\psi_0}\ket{\psi(t)}\approx\nonumber\\
  &&\abs{\exp{- c_0 \rho V_0^{3/\delta} \ev{\abs{\int_0^t s(t')f(t')\dd{t'}}^{3/\delta}}_{s(t)}}}^2\nonumber\\
  &&\sim\exp{- \ev{V_{\delta}}_{\rm typ}^{3/\delta} G_{\delta}(t)},
  \label{eq:F_final}
\end{eqnarray}
where $c_0$ is an adimensional $O(1)$ constant that depends on $\delta$ and on the angular part of the integral, which we absorb in $\ev{V_{\delta}}_{\rm typ}$. This result was given in Eq.~\eqref{eq:fid_decay}.
\subsection{Calculation of $G_{\delta}(t)$}
The function $G_{\delta}(t)$ governing the temporal decay of fidelity can be explicitly rewritten as~\cite{PhysRevB.9.1} a sum over the number $n$ of flips occurring during in the time window $[0,t]$,
\begin{eqnarray}
  G_{\delta}(t)=&&\sum_{n=0}^\infty e^{-\kappa_{s} t}\kappa_{s}^n\prod_{i=1}^n\int_{t_{i-1}}^t\dd{t_n}\abs{\int_0^t s(t';t_1,\dots t_n)f(t')\dd{t'}}^{3/\delta}\nonumber\\
  =&&\sum_{n=0}^\infty g_{n,\delta}(t)
  \label{eq:G_full}
\end{eqnarray}
with $t_0=0$ and 
\begin{equation}
  s(t;t_1,\dots t_n)=s(0)\sum_{j=0}^n(-1)^j\left[\Theta(t-t_j)-\Theta(t-t_{j+1})\right].
\end{equation} 
Rather than evaluating $G_{\delta}(t)$ for all $t$, we examine certain limits. 
We evaluate $G_{\delta}(t)$ at times $t$ that are integer multiples of $\tau$ ($t=N_p\tau$). For a large number of pulses,  $N_p\gg 1$, one finds $3$ regimes:

   {At short times,} $\kappa_{s} t\ll 1$, it is rare for the fluctuator to flip. We may thus just compute the terms $n=0,1$ of the series in Eq.~\eqref{eq:G_full}
  \begin{eqnarray}
    g_{0,\delta}(t)&&=0;\\
    g_{1,\delta}(t)&&=e^{-\kappa_{s} t}\kappa_{s}\int_{0}^t\dd{t_1}\abs{\int_0^{t_1}f(t')\dd{t'} - \int_{t_1}^tf(t')\dd{t'}}^{3/\delta}.\nonumber\\
  \end{eqnarray}
  This results in $G_{\delta}(t)\approx \kappa_{s} t\frac{\delta}{\delta+3}(2\tau)^{3/\delta}=\kappa_{s} t^{1+3/\delta}\frac{\delta}{\delta+3}\left(\frac{2}{N_p}\right)^{3/\delta}$, and leads to stretched exponential decay (with exponent $1+3/\delta$) of the fidelity, with the characteristic time 
  \begin{equation}
      T_{\rm short}^{-1}\sim \kappa_{s}^{\delta/(3+\delta)}\left(\frac{\ev{V_{\delta}}_{\rm typ}}{N_p}\right)^{3/(3+\delta)}
      \label{eq:T_short}
  \end{equation}
  of the short time regime.
  {We note that the weak coupling assumption, $\kappa_s \gg \ev{V_\delta}_{\rm typ}$, implies that the coherence decays less than e-fold within the short time regime, and thus $T_2$ is reached only in later regimes. Indeed, one would need to have $\kappa_s T_{\rm short} < 1$, which, by using Eq.~\eqref{eq:T_short}, requires $\kappa_s < \ev{V_\delta}_{\rm typ}$, in contradiction to our assumption.}
  
  We now discuss the long-time regime, $\kappa_s t\gg 1$, which will depend on the refocusing protocol.
  
   Indeed, if $N_p\gg \kappa_{s} t\gg 1$ (meaning $\kappa_{s}\tau\ll 1$), there are typically multiple fluctuator flips, but rarely more than one per pulse interval. It is then much simpler to evaluate the function as
  \begin{equation}
    G_{\delta}(t)=\ev{\abs{\sum_{p=1}^{N_p}(-1)^p\int_{(p-1)\tau}^{p\tau}s(t')\dd{t'}}^{3/\delta}}_{s(t)}.
  \end{equation}
  If there were no random fluctuator flips, the sum in the last equation would equal zero. Hence, the terms that render the sum non-zero are those corresponding to pulse intervals during which a random flip happens. Thus, $\sum_{p=1}^{N_p}(-1)^p\int_{(p-1)\tau}^{p\tau}s(t')\dd{t'}\approx\sum_{j=1}^{n}(-1)^{p_n}\int_{(p_n-1)\tau}^{p_n\tau}s(t')\dd{t'}$, where $n$ is the number of random flips and $p_n$ labels the pulse intervals during which a random flip has happened. In the limit of large $n$, we can treat this as a sum over $n$ random variables $x_{p_n}$ with average $\ev{x_{p_n}}=0$ and variance $\text{Var}\left[x_{p_n}\right]\sim O(\tau^2)$. 
  Owing to the central limit theorem (CLT), we capture the scaling of the function $G_{\delta}(t)$ by
  \begin{equation}
    G_{\delta}(t)\sim \left(\kappa_{s} t\right)^{\frac{3}{2\delta}}\tau^{\frac{3}{\delta}}=\left(\frac{t^3\kappa_{s}}{N_p^2}\right)^{\frac{3}{2\delta}},
  \end{equation}
  where we have substituted $n$ by its expectation value $n\to\kappa_{s} t$. 

  This implies stretched exponential decay of the fidelity $\mathcal{F}\sim\exp{-\left(\frac{t}{T_{N_p}}\right)^{\frac{9}{2\delta}}}$, with (cf.~Eq.~\eqref{eq:fid_T_N})
 \begin{equation}
 \label{TNp}
 T_{N_p}^{-1}\sim\left(\frac{\ev{V_{\delta}}_{\rm typ}^2 \kappa_{s}}{N_p^2}\right)^{1/3}.
 \end{equation}

   Lastly, if $\kappa_{s} t\gg N_p$ (meaning also $\kappa_{s}\tau\gg 1$), the refocusing technique is ineffective and one may disregard the variation of $f(t)$. Thus, we write
  \begin{equation}
    g_{n,\delta}(t)=e^{-\kappa_{s} t}\kappa_{s}^n\prod_{i=1}^n\int_{t_{i-1}}^t\dd{t_n}\abs{\sum_{j=1}^n(t_j-t_{j-1})(-1)^j}^{3/\delta}.
  \end{equation}
  Again, for large $n$ the sum can be approximated via the CLT, treating the argument of the series as a random variable $x_j$ with zero mean and variance $\text{Var}\left[x_j\right]\sim O\left(\frac{t^2}{n^2}\right)$. Thus,
  \begin{equation}
    g_{n,\delta}(t)\sim e^{-\kappa_{s} t}\frac{(\kappa_{s} t)^n}{n!}\frac{t^{\frac{3}{\delta}}}{n^{\frac{3}{2\delta}}},
  \end{equation}
  which upon summation over  $n$ {using a saddle point approximation} leads to
  \begin{equation}
    G_{\delta}(t)\sim \left(\frac{t}{\kappa_{s}}\right)^{\frac{3}{2\delta}}.
  \end{equation}
Once inserted into Eq.~\eqref{eq:fid_decay}, this yields a different stretched exponential decay with characteristic time scale (cf.~Eq.~\eqref{eq:fid_T_mn})
\begin{equation}
\label{Tmn}
T_{\text{m.n.}}^{-1}\sim\frac{\ev{V_{\delta}}_{\rm typ}^2}{\kappa_{s}}.
\end{equation}

\subsection{Decoherence time $T_2$}
As noted above, e-fold decay of coherence occurs in the intermediate or long time regime. If $\kappa_s/\ev{V_{\delta}}_{\rm typ} \lesssim N_p^{1/2}$, $T_2$ is given by $T_{N_p}$ in \eqref{TNp}, otherwise it is set by $T_{\rm m.n.}$ in \eqref{Tmn}.

\subsection{Higher-order corrections}
So far we have neglected that the interactions from different neighbors do not mutually commute. To assess this approximation, let us thus return to the expansion of the evolution operator of Eq.~\eqref{eq:psi_t}, and keep commutators $[\hat{\phi}_i,\hat{\phi}_j]$, while disregarding higher-order nested commutators, to obtain
\begin{eqnarray}
    &&\hat{\mathcal{T}}\exp{-i\sum_{j\notin\Lambda_{\rm cl}}\int_0^t\hat{\phi}_j(t')\dd{t'}}\approx\prod_{i}\exp\Biggl\{-i\int_0^t{\hat{\phi}_i(t')\dd{t'}}\nonumber\\
    &&\quad\quad\quad\quad -\frac{1}{2}\int_0^t\dd{t_1}\int_{t_1}^t\dd{t_2}\biggl[\hat{\phi}_i(t_1),\sum_{j}\hat{\phi}_j(t_2)\biggr]\Biggr\}.
\end{eqnarray}
Analyzing the corrections arising from the commutators, which we refer to as ``second order" terms, we can check that they either contain the commutator $\left[\hat{V}_i^{\rm dyn}(t_1),\hat{V}_j^{\rm dyn}(t_2)\right]$ or $\left[\hat{V}_i^{\rm dyn}(t_1),\hat{V}_j^{\rm st}(t_2)\right]$ {since at least one factor has to be dynamical in order not to be canceled under the CPMG sequence}. These two commutators contribute differently.

We can estimate the second-order corrections with the commutator of two ``dynamical" parts using the independence of different fluctuators and estimating their commutators as being a randomly signed operator of norm $\sim(\hat{V}^{\rm dyn})^2$.
The contribution from all those commutators will then scale with $t$, the interaction strength and the number of neighbors like the square of the first-order terms considered in Eq.~\eqref{eq:ampl_approx}. {Similar considerations apply to higher-order terms.}

{The specific stretching exponents obtained above in the long-time regime ($\kappa_s t\gg 1$) certainly apply as long as $t < T_2$. At later times, however, higher-order contributions might lead to a crossover into a faster decay law. Thus, the behavior in the tail may be altered, although its characteristic time scale will stay parametrically the same.}

\section{Formal treatment of ring-exchange decay exponent}
\label{sec:supp_ring_exp}
For each term of the multipole expansion of Eq.~\eqref{eq:multipole-exp}, we decompose the representations $\Gamma_n$ into irreps, $\Gamma_n=\bigoplus_i\Gamma_n^{(i)}$.
Note that as a consequence of time reversal invariance, the Hamiltonian is real. If the character table for the point group $G_{\rm cl}$ contains complex irreps, they can thus be grouped into real representations. {Below we therefore assume the representations to be real, satisfying $\Gamma^*=\Gamma$.}
 
To obtain a term with exponent $\xi$, there needs to be at least one pair of irreps $(\Gamma_n^{(i)},\Gamma_m^{(j)})$ with $n+m=\xi$, such that the following conditions are satisfied:

\begin{enumerate}
  \item The ring-exchange interaction has non-zero matrix elements in the $\Gamma_d$ subspace. Because it is the second-order term of the perturbative expansion of $J(\bm{r})$, this is equivalent to the condition that the trivial representation is contained in the following tensor product
  \begin{equation}
  A\subseteq\Gamma_d\otimes\Gamma_n^{(i)}\otimes\Gamma_m^{(j)}\otimes\Gamma_d
  \end{equation}
  \item There exists at least one multiplet with irrep $\Gamma'$ in the reachable manifolds with $M\pm 1$ excitations, for which
  \begin{eqnarray}
  A\subseteq\Gamma_d\otimes\Gamma_n^{(i)}\otimes\Gamma',\\
  A\subseteq\Gamma_d\otimes\Gamma_m^{(j)}\otimes\Gamma'.
  \end{eqnarray}
  These conditions express the possibility of a virtual exchange of an excitation via an intermediate state with irrep $\Gamma'$.

  \item Finally, the considered order of the ring-exchange has to split the doublet degeneracy. This requires that the action of the two two multipole terms does not reduce to the mere multiplication by a number,
  \begin{equation}
    \Gamma_n^{(i)}\otimes\Gamma_m^{(j)}\neq A.
  \end{equation}

Large symmetry suppression exponents $\xi>2$ are unlikely to occur. {Indeed, already $\xi=3$ would require that there be no coupling at the dipole order to any irrep $\Gamma'$ in the manifolds $M'=M\pm 1$; or mathematically: for all terms $\Gamma_2^{(i)}$, $A\not\subset\Gamma_d\otimes\Gamma_2^{(i)}\otimes\Gamma'$. We have not found any such case in simple clusters.}

\end{enumerate}

\section{Examples of effective interaction and ring-exchange}
\label{sec:supp_examples}
In this Appendix, we analyze two instructive cases in more detail: an $S_4$-symmetric cluster and a $C_4$-symmetric cluster, {both containing $N_{\rm cl}=4$ sites}.  
In the end, we briefly comment on $C_3$-symmetric clusters of 3 sites.
We use Mulliken's notation for the specific irreps of these two groups. 

\begin{figure}[ht!]
  \centering
  \includegraphics[width=0.48\textwidth]{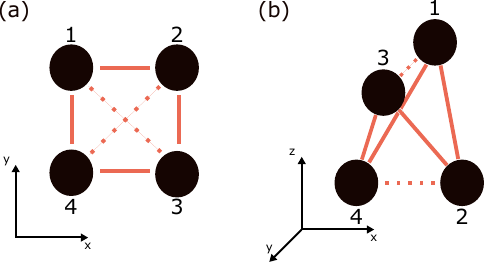}
   \caption{Pictorial representation of the (a) $C_4$ and (b) $S_4$ clusters, with the labeling used in Eq.~\eqref{eq:H_4}. A solid line represents an interaction $J_a$, while a dashed line corresponds to an interaction $J_b$.}
  \label{fig:fig_supp}
\end{figure}

For both the $S_4$ and $C_4$ symmetric cases, we write the cluster Hamiltonian as in Eq.~\eqref{eq:H_cl}, with the site labeling shown in Fig.~\ref{fig:fig_supp}.
\begin{eqnarray}
  \hat{H}_{4}&&=\frac{\Delta}{2}\sum_{i=1}^{4} \hat{\sigma}_i^z + J_a\sum_{i=1}^4{\left(\hat{\sigma}^+_i\hat{\sigma}^-_{i+1} + \hat{\sigma}^-_i\hat{\sigma}^+_{i+1}\right)} + \nonumber\\ &&+J_b\sum_{i=1,2}{\left(\hat{\sigma}^+_i\hat{\sigma}^-_{i+2} + \hat{\sigma}^-_i\hat{\sigma}^+_{i+2}\right)},
  \label{eq:H_4}
\end{eqnarray}
with $J_a$ and $J_b$ being two different interaction strengths {(identifying $\bm{\sigma}_{5}=\bm{\sigma}_1$)}. The ground state has energy $\epsilon_0=-2\Delta$, and in both cases the $M=1$ and $M=2$ manifolds each host a symmetry-protected doublet, with energies $\epsilon_{1,E}=-\Delta + J_a$ and $\epsilon_{2,E}=0$.
The $M=1$ doublet states can be written as two ``spin-singlet" states involving either odd or even ions
\begin{eqnarray}
   &&\ket{1,E,1}=\ket{1}_1\ket{0}_2\ket{0}_3\ket{0}_4-\ket{0}_1\ket{0}_2\ket{1}_3\ket{0}_4,\nonumber\\&&\ket{1,E,2}=\ket{0}_1\ket{1}_2\ket{0}_3\ket{0}_4-\ket{0}_1\ket{0}_2\ket{0}_3\ket{1}_4,
\end{eqnarray}
while for $M=2$
\begin{eqnarray}
  &&\ket{2,E,1}=\ket{1}_1\ket{1}_2\ket{0}_3\ket{0}_4-\ket{0}_1\ket{0}_2\ket{1}_3\ket{1}_4,\nonumber\\&&
  \ket{2,E,2}=\ket{1}_1\ket{0}_2\ket{0}_3\ket{1}_4-\ket{0}_1\ket{1}_2\ket{1}_3\ket{0}_4,
\end{eqnarray}
(see Fig.~\ref{fig:fig2}). From the point of view of the doublet wavefunctions, the two clusters are thus identical.
However, what differentiates the two cluster symmetries is the geometric arrangement of the ions and, as a result, the coupling to an external ion. {Mathematically speaking it is the representations of the multipole terms, $\Gamma_n$, which differ for the two clusters.}

For the $S_4$-symmetric cluster, the {monopole and dipole orders of the interaction with an external ion, cf.~Eq.~\eqref{eq:multipole-exp}, are associated respectively with the representations $\Gamma_0=A\otimes A=A$ and $\Gamma_1=A\otimes (B\oplus E)=B\oplus E$. Note that the dipolar representation contains a $B$ irrep, since the $z$ coordinate of the ions changes sign under the action of the $S_4$ generator, implying $\Gamma_z=B$}. The $B$ irrep induces non-trivial transitions between the $E$ doublet and the doublet in the $M=2$ manifold, {which can be connected back to the $M=1$ doublet via the (trivial) monopole term.} Their combination, $B\otimes A$, thus acts non-trivially on the doublet (satisfying the third condition of App.~\ref{sec:supp_ring_exp}). The doublet degeneracy is thus lifted at first order in the multipole expansion, and the ring-exchange exponent is $\xi=1$ for this cluster (cf.~Sec.~\ref{subsec:ring_ex}). 
Moreover, the $E$ irrep connects the $M=1$ doublet with the ground state, and thus the effective interaction exponent, {which governs relaxation through excitation hopping,} also takes the value $\nu =1$. 

A $C_4$ cluster instead behaves differently. The corresponding representation for the {first two multipole terms} in the perturbing interaction are $\Gamma_0=A$ and $\Gamma_1=A\otimes(A\oplus E)=A\oplus E$, since here {the $z$ coordinate of the ions are all identical, giving rise to the component $A$ in $\Gamma_{\bm{r}}$.} Hence, this term connects doublets only trivially via the irrep $A$. Therefore $\xi=2$, {since the quadrupolar term contains a $B$ irrep}.

Let us finally discuss a smaller cluster of $3$ ions with a $C_3$ symmetry. One finds the same characteristic exponents $\xi=1$ and $\nu=1$ as for the $S_4$-symmetric 4-site cluster. However, the transition {frequency} between the $M=1$ and $M=2$ doublets turns out to be independent of the interaction strength $J$ {and identical to the single ion CF splitting}. Hence, the coupling to neighboring ions, even though symmetry-reduced, is not spectrally suppressed, but rather resonant - unless the CF energy shift induced by the cluster ions is sizeable~\cite{fricke1979satellite,PhysRevLett.117.037203,hu2022single}. If this is not the case, resonant excitation hopping is likely to be the leading dephasing mechanism which would mask the weaker ring-exchange contribution.

\bibliography{apssamp}% Produces the bibliography via BibTeX.

\end{document}